\documentclass[aps,preprint,amsmath,amssymb,amsfonts,nofootinbib]{revtex4}
\usepackage{epsfig}
\usepackage{graphicx}
\usepackage{dcolumn}
\usepackage{bm}
\usepackage{amsthm}
\usepackage{amsmath}
\usepackage{color}

\newcommand{\beq}{\begin{equation}}
\newcommand{\eeq}{\end{equation}}
\newcommand{\bea}{\begin{eqnarray}}
\newcommand{\eea}{\end{eqnarray}}

\begin{document}

\title{New Asymptotically Lifshitz Black Holes in Ho\v{r}ava gravity}
\author{Christopher Eling}
\email{cteling@gmail.com}
\affiliation{Rudolf Peierls Centre for Theoretical Physics, University of Oxford, 1 Keble Road, Oxford OX1 3NP, UK}

\begin{abstract}

We study asymptotically Lifshitz solutions with critical exponent $z \neq 1$ in Ho\v{r}ava gravity in three and four spacetime dimensions. For $z=2$ and $z=3/2$, we find a novel class of numerical solutions with regular universal horizon, but are characterized by non-analytic behavior near infinity. In the interior, inside the universal horizon, the unit timelike vector field associated with the preferred time foliation exhibits oscillatory behavior, qualitatively similar to that found earlier in asymptotically flat solutions. For $z>2$ no solutions of this type appear to exist. We comment on potential applications to holographic Lifshitz dualities.

\end{abstract}

\pacs{...}

\maketitle

\section{Introduction}

Over the past decade holography has become an important tool to describe strongly coupled condensed matter systems. Since many real world systems of interest are non-relativistic, there has been growing interest in constructing gravitational models that exhibit non-relativistic Galilean or Lifshitz symmetries. In this paper we will focus on dual Lifshitz theories, where boost invariance is broken and space and time scale in an anisotropic way $x^i \rightarrow \lambda x^i, ~ t \rightarrow \lambda^z t$. The dynamical exponent $z$ in general will differ from unity. A field theory with this type of symmetry is thought to describe quantum critical points, which characterize zero temperature phase transitions associated with quantum fluctuations \cite{subir}.

On the gravitational side, it was shown that Einstein gravity with various additional matter fields in the bulk can have solutions with asymptotic Lifshitz scaling \cite{Kachru:2008yh,Taylor:2008tg,Goldstein:2009cv}. In these cases the dual gravitational theory is of course relativistic, with local Lorentz invariance. The Lifshitz scaling emerges only as feature of a special class of solutions, supported by the exotic matter fields.  A more natural, alternative dual description is in terms of a gravity theory with a preferred frame \cite{Griffin:2011xs,Janiszewski:2012nb}.

One such theory is Einstein-aether theory \cite{AEtheory}, where in addition to the metric there is a dynamical unit timelike vector field acting as an ``aether". Einstein-aether theory has been studied extensively as an alternative theory of gravity (i.e. focusing on asymptotically flat and cosmological solutions), for a review see \cite{AEreview}.  Another theory is Ho\v{r}ava-Lifshitz gravity, or Ho\v{r}ava gravity for short, which has attracted much attention as a possibly renormalizable quantum theory of gravity \cite{Horava:2009uw}. Ho\v{r}ava gravity can be thought of in the low energy limit as Einstein-aether gravity restricted to the case where the aether is hypersurface orthogonal, defining a preferred time.

In \cite{Griffin:2012qx} the authors showed that Ho\v{r}ava gravity (with negative cosmological constant) is a viable setting for non-relativistic Lifshitz holography and worked out some aspects of the holographic dictionary.  In holography, thermal states in the field theory are dual to black hole solutions in the bulk gravitational theory.  Therefore to study Lifshitz hydrodynamics we need to find black hole solutions and their associated thermodynamics. However, in a theory of gravity with a preferred frame the loss of relativistic causality leads to the existence of apparently multiple horizons, which clashes with the laws of black hole thermodynamics see e.g. \cite{Dubovsky:2006vk, Eling:2007qd, Jacobson:2008yc}. A potential resolution is to define a black hole via the existence of a ``universal horizon", a spacelike surface where even modes of arbitrary speed do not reach infinity \cite{Barausse:2011pu,Blas:2011ni}. The universal horizon is defined in terms of the preferred time foliation, in particular it is the surface where the preferred time $\tau \rightarrow \infty$. Several works have shown that one can indeed define a notion of temperature and a First law for the universal horizon \cite{Berglund:2012bu,Berglund:2012fk,Mohd:2013zca, Cropp:2013sea}.

Asymptotically Lifshitz black holes with regular universal horizons have previously been found in Ho\v{r}ava gravity. In particular, in the special case where $z=1$ there is an analytic black hole solution \cite{Janiszewski:2014iaa} in four dimensions. One can perturb this solution in the context of the fluid-gravity correspondence and read off some of the transport coefficients of the dual theory \cite{Eling:2014saa, Davison:2016auk}. However, the generic $z \neq 1$ cases are of the most interest. Recently, \cite{Basu:2016vyz} found a numerical solution in three dimensions for $z=2$ and special choice of coupling constants.

In this paper we show numerically that there exist a wide class of solutions with regular universal horizon for $z=2$ and $z=3/2$ in both three and four dimensions. Near infinity, these solutions have non-analytic behavior, where series expansions are in terms of non-integer powers. We also explore the interior and show there is oscillatory behavior in the aether field, just as in asymptotically flat cases. Curiously, for $z>2$ no solutions of this type appear to exist. The structure of this paper is as follows. In Section II we briefly review Einstein-aether theory and how hypersurface orthogonal solutions are also solutions of Ho\v{r}ava gravity. In Section III we discuss asymptotic Lifshitz solutions, which can have analytic and non-analytic behavior when $z \neq 1$.  In Section IV, we discuss generic features of black hole solutions, including how one imposes regularity. In Section V we present our results. We conclude with a discussion and areas for future research.

\section{Einstein-aether and Ho\v{r}ava-Lifshitz gravity}

The action for Einstein-aether theory with a negative cosmological constant $\Lambda$ is
\begin{align}
S_{ae} = \frac{1}{16\pi G_{ae}} \int d^{D+1} x \sqrt{-g} L_{ae} \label{totalaction} \ ,
\end{align}
where $L_{ae} = R + L_{vec} + 2 \Lambda$,  with
\begin{align}
-L_{vec} = K^{AB}{}_{CD} \nabla_A v^C \nabla_B v^D - \lambda(v^2+1) \ ,
\end{align}
and
\begin{align}
K^{AB}{}_{CD} = c_1 g^{AB} g_{CD} + c_2 \delta^A_C \delta^B_D + c_3 \delta^A_D \delta^B_C - c_4 v^A v^B g_{CD} \ .
\end{align}
This turns out to be the most general effective action for a timelike unit vector field at 2nd order in derivatives (a term proportional to Ricci tensor contracted with two aether vectors is a combination of the $c_2$ and $c_3$ terms above).

Varying this action with respect to the metric, vector field, and the Lagrange multiplier $\lambda$, one finds the following field equations
\begin{align}
G_{AB} + \Lambda g_{AB} = T^{ae}_{AB}, ~~ E_A = 0, ~~ v^2 = -1 \ .
\end{align}
The aether stress tensor is given by
\begin{align}
T^{ae}_{AB} = \lambda v_A v_B + c_4 a^{(v)}_A a^{(v)}_B - \frac{1}{2} g_{AB} Y^C{}_D \nabla_C v^D + \nabla_C X^C{}_{AB} + c_1[(\nabla_A v_C)(\nabla_B v^C)-(\nabla^C v_A)(\nabla_C v_B)] \ ,
\end{align}
where
\begin{align}
Y^A{}_B =& K^{AC}{}_{BD} \nabla_C v^D  \ , \\
X^C{}_{AB} =& Y^C{}_{(A} v_{B)} - v_{(A} Y_{B)}{}^C + v^C Y_{(AB)} \ ,  \
\end{align}
and $a^{(v)}_A = v^B \nabla_B v_A$. The aether field equation is
\begin{align}
E_A = \nabla_B Y^B{}_A + \lambda v_A + c_4 (\nabla_A v^B) a^{(v)}_B \label{aefieldeqn} \ .
\end{align}

One can show that solutions of Einstein-aether theory where the aether is hypersurface orthogonal are also solutions of Ho\v{r}ava-Lifshitz gravity \cite{Jacobson:2010mx}. In this case the aether foliation defines a preferred time foliation in the spacetime.  First note that when the aether field is hypersurface orthogonal, the twist
\begin{align}
\omega_{AB} = v_{[A} \nabla_B v_{C]} \
\end{align}
vanishes. One can show that
\begin{align}
\omega^2 = (\nabla_A v_B)(\nabla^A v^B)-(\nabla_A v_B)(\nabla^B v^A)+a_{(v)}^2 \ .
\end{align}
Since squared twist must also vanish, adding it to the action doesn't affect the solutions. Due to the fact that this scalar twist is a combination of the $c_1$, $c_3$ and $c_4$ terms, we can use the freedom to add $\omega^2$ to eliminate one of $c_1$, $c_3$ or $c_4$ in the action. A convenient choice is to eliminate $c_1$, therefore we take $c_1=0$ from this point forward.

Secondly, hypersurface orthogonality implies the co-vector is the gradient of a scalar
\begin{align}
v_A = \frac{-\partial_A \phi}{\sqrt{g^{CD} \partial_C  \phi \partial_D \phi}} \ .
\end{align}
If we choose coordinates such that $\phi=\tau$, where $\tau$ is the preferred foliation of time, then in this gauge the Einstein-aether action reduces to the generic $D+1$ form of the Ho\v{r}ava-Lifshitz action (e.g. \cite{Griffin:2012qx})
\begin{align}
S_{HL} = \frac{1}{16\pi G_H} \int d\tau d^3 x \sqrt{\gamma} \left(K_{ab} K^{ab} - (1-\lambda) K^2 +(1+\beta) R^{(3)} + \tilde{\alpha} \frac{\nabla_a N \nabla^a N}{N^2} \right)
\ \label{HLaction}.
\end{align}
Here $K_{ab}$ is the extrinsic curvature of the preferred time slices, $\gamma_{ab}$ the spatial metric on the slices, $R^{(3)}$ the intrinsic Ricci scalar and $N$ is the lapse function, $v_A = -N \delta^\tau_A$. The mapping between the constants is given by
\begin{align}
1+\lambda = \frac{1+c_2}{1-c_3}, ~~ \tilde{\alpha} = \frac{c_4}{1-c_3}, ~~ \frac{G_H}{G_{ae}} = 1+\beta = \frac{1}{1-c_3} \ .
\end{align}
In Einstein-aether theory there are five propagating degrees of freedom with spin-2, spin-1, and spin-0 helicities \cite{Jacobson:2004ts}. In Ho\v{r}ava-Lifshitz the spin-1 mode is non-propagating. The squared speeds of the remaining modes are given in general space dimension $D$ by \cite{Griffin:2012qx}
\begin{align}
s_2^2 = \frac{1}{1-c_3}, ~~ s_0^2 = \frac{(c_2+c_3)(D-1-c_4(D-2))}{c_4(1-c_3)(D-1 + D c_2+c_3)} \label{speeds} \ .
\end{align}

\section{Lifshitz Asymptotic Solutions}

We begin to characterize solutions by considering the following ansatz for the metric and aether in terms of Eddington-Finkelstein like coordinates
\begin{align}
ds^2 = F(\rho) dt^2 - 2 G(\rho) dt d\rho + \rho^2 dx_i dx^i, \\
v_A dx^A = \frac{G(\rho)^2-F(\rho)K(\rho)^2}{2 K(\rho) G(\rho)} dt + K(\rho) d\rho \ .
\label{ansatz}
\end{align}
Here the aether field is hypersurface orthogonal (vanishing twist), so that solutions will satisfy both the Einstein-aether and Ho\v{r}ava-Lifshitz field equations. In practice, we will work with the Einstein-aether field equations. Plugging the ansatz into these equations leads to a complicated set of second order ordinary differential equations for the functions $F(\rho)$, $G(\rho)$, and $K(\rho)$.  One can take solutions in this case and re-express them in a gauge where $v_A$ only has time component. In this frame, the solution will also satisfy the Ho\v{r}ava-Lifshitz field equations.

In holography, the coordinates $(t,x^i)$ are identified with the dual field theory, while $\rho$ is the bulk radial coordinate. Lifshitz scaling by definition is an anisotropic scaling of space and time in the field theory: $x^i\rightarrow \lambda x^i, t\rightarrow \lambda^z t$.  In the full bulk metric we require an invariance under this scaling, with $\rho \rightarrow \lambda^{-1} \rho$. This leads to the asymptotic form as $\rho \rightarrow \infty$
\begin{align}
ds^2 \sim - \rho^{2z} dt^2 + 2 \rho^{z-1} dt d\rho + \rho^2 dx_i dx^i \nonumber \\
K(\rho) \sim \frac{1}{\rho} \label{asymptotics} \ .
\end{align}
When $z=1$ we have the standard AdS asymptotic metric. If we plug this generic form into the field equations, one finds the following consistency conditions for it to be a solution \cite{Griffin:2012qx},
\begin{align}
c_4 = \frac{z-1}{z}, \Lambda = - \frac{(D-2+z)(D-1+z)}{2} \ .
\end{align}

Asymptotically Lifshitz solutions were first studied in detail in \cite{Janiszewski:2014iaa} in four-dimensional spacetime ($D=3$) by expanding the field equations in a power series around $r = 1/\rho = 0$. It was found that the behavior of solutions separates into two classes depending on if $z=1$ or $z \neq 1$. In the $z=1$ case there is a solution characterized by two free parameters $C_e$ and $C_a$. The series solution for $G(\rho)$ and $F(\rho)$ truncates and it was ultimately shown that there is a full analytic solution for general $\rho$ given by
\begin{align}
G(\rho) =& \rho^2 \nonumber \\
F(\rho) =& \rho^2 + \frac{C_e}{\rho} + \frac{c_3 (C_e+2C_a)^2}{4 \rho^4} \nonumber \\
K(\rho) =& 2\left(\frac{\sqrt{4 \rho^{14} + 4C_e \rho^{11} + (1-c_3)(C_e+2C_a)^2 \rho^8} + (C_e+2C_a)\rho^4}{4\rho^5(\rho^3+C_e)-c_3(C_e+2C_a)^2 \rho^2}\right). \label{analytic}
\end{align}
This solution depends only on $c_3$. Note that for $z=1$, $c_4 =0$.

In the $z \neq 1$ case there appears to be no analytic solution for general $\rho$. The behavior of asymptotic solutions was determined to be dependent on the analyticity properties of the solution ansatz. If we assume the functions $F(\rho)$, $G(\rho)$, and $K(\rho)$ are of the form of their leading asymptotics (\ref{asymptotics}) times an analytic function, then the resulting series solution depends generically only on one free parameter. On general grounds \cite{Eling:2006ec} there should be a two parameter family of solutions near infinity. This indicates this ansatz is a restrictive special case that does not reflect the general properties of the asymptotic solutions.

Instead \cite{Janiszewski:2014iaa} noted that one can allow for non-analytic behavior by imposing
\begin{align}
F(\rho) =& \rho^{2z} \nonumber \\
G(\rho) =& \rho^{z-1} \nonumber \\
K(\rho) =& \rho^{-1}(1+a_{\Delta} \rho^{-\Delta}).
\end{align}
Substituting into the field equations, it was found that
\begin{align}
\Delta = \frac{1}{2} \left(z+2 \pm \sqrt{(z+2)^2 - \frac{8(1-c_3)(z-1)}{c_2+c_3}} \right), \label{Deltacondition}
\end{align}
restricting attention to non-negative cases. In general $\Delta$ is not an integer.  In this way it was found that $z \neq 1$ solutions are indeed characterized generically by two free parameters which are the coefficient of $\rho^{-\Delta-1}$ in the expansion for $K(\rho)$ and the coefficient of $\rho^{z-2}$ in the expansion of $F(\rho)$.  Note that in special cases where $z \neq 1$ one can choose $c_2$ so that $\Delta$ in (\ref{Deltacondition}) is an integer. In this particular case one has an analytic power series solution around infinity that depends on two free parameters.

Recently, \cite{Basu:2016vyz} studied asymptotic Lifshitz solutions for the three-dimensional case ($D=2$). They imposed the following condition of analyticity on the power series solutions around infinity
\begin{align}
F_1(\rho) = \left(1+ \frac{f_1}{\rho^{\nu_\star}} +  \frac{f_2}{\rho^{2\nu_\star}} + \cdots \right) \nonumber \\
G_1(\rho) =  \left(1+ \frac{g_1}{\rho^{\nu_\star}} +  \frac{g_2}{\rho^{2\nu_\star}} + \cdots  \right)  \nonumber \\
-U(\rho) = \rho^{z-1} \left(1+ \frac{u_1}{\rho^{\nu_\star}} +  \frac{u_2}{\rho^{2\nu_\star}} + \cdots  \right)   \label{mattingly},
\end{align}
where $U(\rho) = v_A \xi^A$ and $\xi^A$ is the timelike Killing vector field. In order to have a non-zero and finite ADM-like mass it was found that  $\nu_\star = z+1$. The field equations imply that there is a restriction to a subset of solutions with particular $c_2$. The relevant condition on the couplings is
\begin{align}
c_2 + c_3 = \frac{4(1-c_3)(z-1)}{n_s (n_s-2) (z+1)^2},  \label{3dcondition}
\end{align}
where $n_s$ is an integer $\geq 3$. The solution again depends on two free parameters. The advantage of these power series solutions is that one can define and read-off the total mass (per unit volume) of the black hole and then construct a First Law of thermodynamics based on the temperature and entropy of a universal horizon. However, in the following we will show in many cases that there are regular black holes with universal horizons that do not necessarily have analytic behavior near infinity.

\section{Regular Black Hole Solutions}

Asymptotically Lifshitz solutions are generally characterized by two free parameters. However, there is no guarantee that these solutions are regular everywhere in the bulk interior. Singularities generically arise at the location of the horizon associated with the spin-0 mode \cite{Eling:2006ec,Barausse:2011pu}. In the $z=1$ case, the speed of the spin-0 mode (\ref{speeds}) is infinite, so the horizon in this case coincides with the universal horizon. The universal horizon is defined as the spacelike surface where $\xi^A$  is orthogonal to the aether, i.e. $U(\rho) = 0$. For the $D=3$ global analytic solution, imposing regularity at the universal horizon amounts to one additional condition relating $C_e$ to $C_a$, \cite{Janiszewski:2014iaa} i.e.
\begin{align}
C_e = - 2 \rho_h^3, ~~C_a = \rho_h^3 \left(1-1/\sqrt{1-c_3}\right),
\end{align}
where $\rho_h$ is the location of the universal horizon. $\rho_h^{3}$ is proportional to the total mass per area of the black hole. Therefore one is left with parameter family of regular black hole solutions. An analogous solution in the $D=2$ case was later found in \cite{Sotiriou:2014gna}.

We now want to determine numerically whether other one parameter families of regular solutions exist when $z \neq 1$. Following  \cite{Eling:2006ec,Barausse:2011pu} the idea is to impose regularity at the spin-0 horizon and then integrate outwards to infinity, demanding the Lifshitz asymptotic conditions (\ref{asymptotics}). Practically, to implement this we redefine
\begin{align}
F(\rho) =& \rho^{2z} F_1(\rho) \nonumber \\
G(\rho) =& \rho^{z-1} G_1(\rho) \nonumber \\
K(\rho) =& \rho^{-1} K_1(\rho)
\end{align}
where $F_1(\rho)$ and $G_1(\rho)$ should approach a constant plus corrections near infinity, while $K_1$ goes to unity. Asymptotic values of $F_1$ and $G_1$ depend on a choice of initial conditions at the spin-0 horizon, which is a reflection of the scaling freedom in the time coordinate $t \rightarrow \alpha t$. Under this rescaling
\begin{align}
F_1 \rightarrow \alpha^2 F_1 \nonumber \\
G_1 \rightarrow \alpha G_1 \nonumber \\
B_1 \equiv  \frac{K_1}{G_1} \rightarrow \alpha B_1,
\end{align}
but the combinations $N_1 = F_1 B_1^2$ and $M_1 = G_1 B_1$ are invariant. Thus asymptotically Lifshitz conditions are imposed universally by demanding these combinations of functions approach unity at infinity.

For simplicity, the next step is to fix $s_0=1$ so that the location of the spin-0 horizon coincides with the usual metric/Killing horizon in the bulk which is where $F(\rho) = 0$. Fixing $z$ fixes $c_4$ and then this condition can be used to solve for $c_2$
\begin{align}
c_2 = \frac{1-D+D z - z + c_3 z -c_3 D z + c_3^2 - c_3^2 z}{z-2+2D-c_3 D - D z +c_3 D z}.  \label{c2condition}
\end{align}
This choice can be made without loss of generality by making the following field redefinition of the metric and the aether for a theory with arbitrary $c_i$
\begin{align}
\bar{g}_{AB} =& g_{AB} + (s_0^2 - 1) v_A v_B \nonumber \\
\bar{v}^A =& \frac{1}{s_0} v^A,
\end{align}
which maps $\bar{c}_i$ into $c_i$ and has the effect of mapping arbitrary $s_0$ into $s_0 = 1$.

To impose regularity at the horizon we consider a series solution ansatz
\begin{align}
F_1(\rho) =& F_1'(\rho_0) (\rho-\rho_0) + \cdots\nonumber \\
G_1(\rho) =& G_1(\rho_0) + G_1'(\rho_0)(\rho-\rho_0) + \cdots \nonumber \\
K_1(\rho) =& K_1(\rho_0) + K_1'(\rho_0)(\rho-\rho_0) + \cdots
\end{align}
where $\rho_0$ is the location of the spin-0 horizon. Using the time rescaling freedom discussed above we can fix one value, say $G_1(\rho_0)$, at the horizon. Plugging into the field equations, we find that the values of $F_1'(\rho_0)$, $G_1'(\rho_0)$ and $K_1'(\rho_0)$ are determined (in a very complicated form using Maple) in terms of $\rho_0$ and $K_1(\rho_0)$. Thus, there is a two parameter family of regular spin-0 horizon solutions in the Lifshitz case, which agrees with the earlier $\Lambda=0$ results in \cite{Eling:2006ec}.  These local solutions of course will not generically be asymptotically Lifshitz.

\section{Asymptotically Lifshitz Black Holes for $z \neq 1$}

To find regular Lifshitz black holes we use the shooting method. We fix a small value of $x=\rho-\rho_0$ and use the series solution above to find the initial radial values. Then we integrate the field equations, a set of ordinary differential equations, out to infinity. The initial values depend on $K_1(\rho_0)$ and we tune this value until the combinations $N_1$ and $M_1$ approach unity at very large values of $\rho$, say $10^8$. If the procedure finds a solution we will have a fixed numerical value of $K_1(\rho)$, meaning that the asymptotic condition has eliminated an additional free parameter. The resulting solution will depend on one free parameter $\rho_0$ (similar to the Schwarzschild solution in GR, asymptotically flat Einstein-aether black holes in \cite{Eling:2006ec,Barausse:2011pu} and finally the $z=1$ solution found in \cite{Janiszewski:2014iaa}). To determine the solution in the interior, we then take the value of $K_1(\rho_0)$ and use the series solution for small negative $\rho-\rho_0$ to provide the necessary initial data to integrate inwards.

\subsection{$D=2$ Black Holes}

First we study the three-dimensional Ho\v{r}ava-Lifshitz theory can focus first on the $z=2$ case. $z =2$ is likely the case of most relevance for physical systems since in field theory the symmetry group can be enhanced to include Galilean symmetries. For a range of $c_3$ up to about $\sim 0.4$ we find regular solutions. In the shooting process one can see that for a range of $K_1(\rho_0)$ the values of $N_1$ and $M_1$ are $\gg 1$ for very large $\rho$ while for others they become $\ll 1$.  The tuning leads to the set of values displayed in Table 1 where the asymptotic conditions are met.
\small
\begin{table}
\caption{\label{bhproperties} Values of the shooting parameter for
several $c_3$ values in $D=2$, in units with $r_0=1$.}
\begin{center}
\begin{tabular}{| | c c | |}
\hline
$c_3$&$K_1(\rho_0)$\\
\hline
  0.1 &  1.06970\\
  \hline
  0.2 &  1.06169\\
  \hline
  0.3 &  1.05304\\
  \hline
  0.33 & 1.05031\\
\hline
\end{tabular}
\end{center}
\end{table}
\normalsize
The behavior of the functions for $c_3=0.1$ is shown in Figure 1. We have chosen the standard normalization/gauge where $F_1$ and $G_1$ approach unity at infinity.
\begin{figure}
\begin{center}
 \includegraphics[angle=0,height= 11cm, width=11cm]{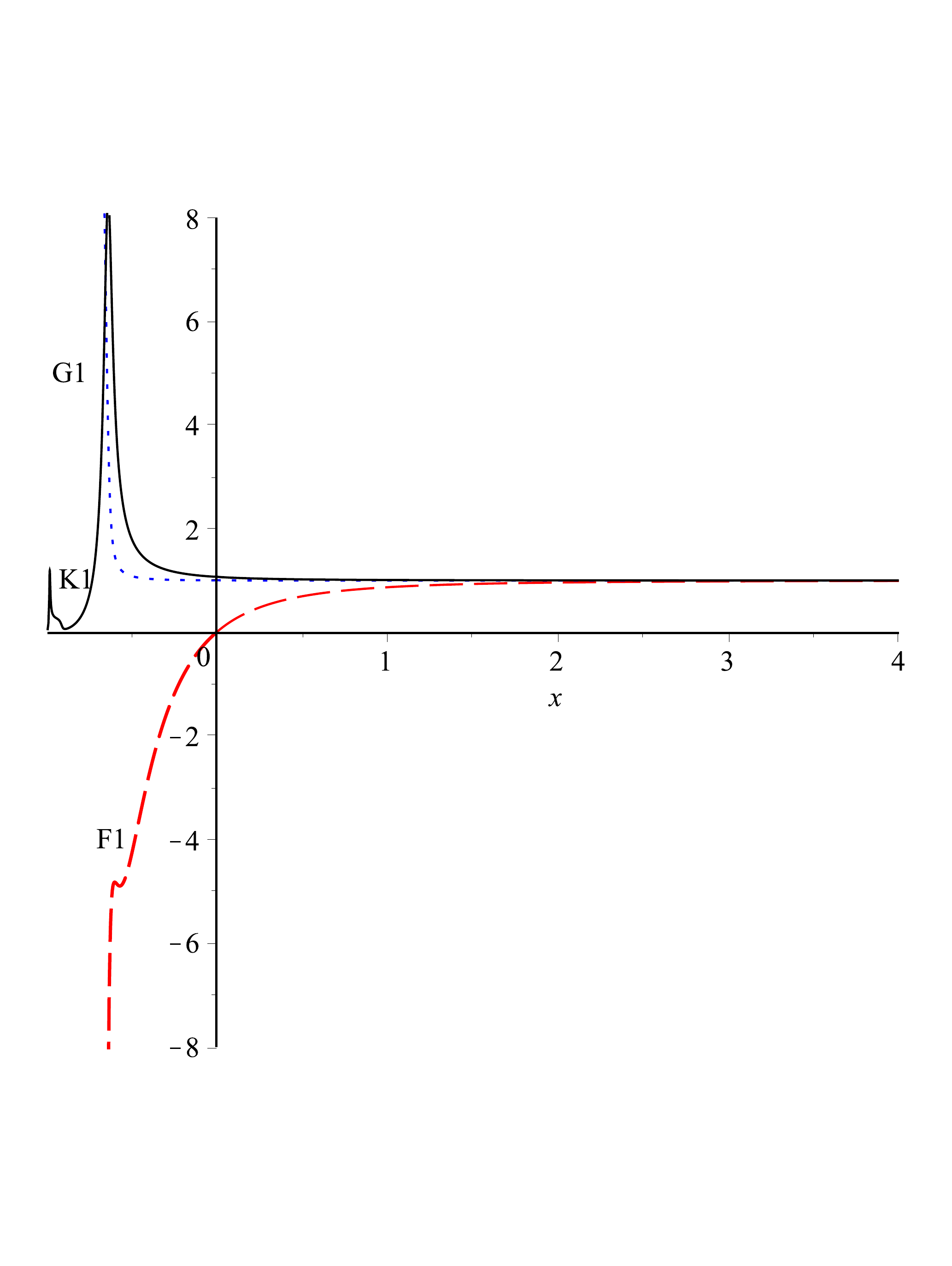}\\
\caption{\label{bhsol1}Plots of $F_1$ (dashed), $G_1$ (dotted) and $K_1$ (solid) vs $x=r-r_0$ for $c_3=0.1$ in $D=2$. Here $r_0$ is chosen to be 1. In the interior the functions $F_1$ and $G_1$ diverge as $x=-1$ consistent with a curvature singularity. }
 \end{center}
\end{figure}
We performed a non-linear curve fit of this data to the asymptotic forms
\begin{align}
F_1(\rho) =& 1 + \frac{f_1}{\rho^{\alpha}} + \frac{f_2}{\rho^{2 \alpha}} \nonumber \\
G_1(\rho) = & 1 + \frac{g_1}{\rho^{\beta}} + \frac{g_2}{\rho^{2 \beta}}  \nonumber \\
K_1(\rho) = & 1+ \frac{k_1}{\rho^{\gamma}} + \frac{k_2}{\rho^{2 \gamma}}
\end{align}
the results yielded $\alpha \sim 3.00$, $\beta \sim 2.28$, $\gamma \sim 1.40$, which differs from the fall-off condition (\ref{mattingly}) in (\cite{Basu:2016vyz})  and is consistent with non-analytic (non-integer expansion) in $K_1(\rho)$ and $G_1(\rho)$.

While the solutions have a regular spin-0 horizon, for there to be a black hole in Ho\v{r}ava-Lifshitz gravity, modes of arbitrary speed must be trapped, so we require a universal horizon. A universal horizon is located where
\begin{align}
U(\rho) = -\frac{FK^2+G^2}{2KG} = -\rho^{z-1} \left(\frac{F_1 K_1^2+G_1^2}{2K_1 G_1} \right)
\end{align}
vanishes. For $x \geq 0$ $U(\rho)$ is negative.  A plot of $U(\rho)/\rho$ versus $x$ in the interior is shown in Figure 2.
\begin{figure}
\begin{center}
 \includegraphics[angle=0,height= 11cm, width=11cm]{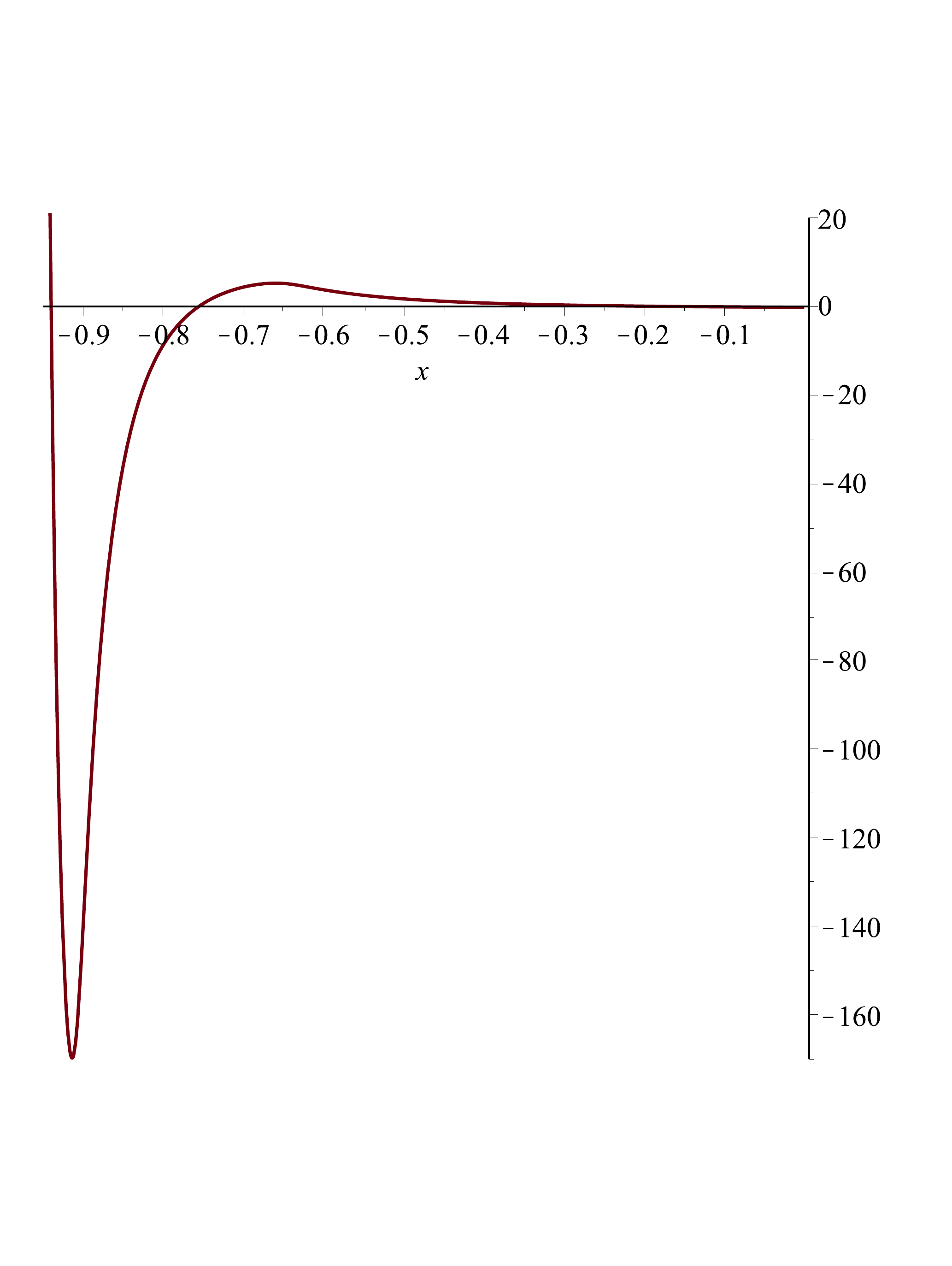}\\
\caption{\label{universal}Plot of $U(\rho)/\rho$ versus $x$ for the solution in Figure 1. The function has multiple roots where universal horizons are present and begins to oscillate wildly as the singularity is approached. }
 \end{center}
\end{figure}
At $x \sim -0.1887$, $U(\rho)$ vanishes so there is indeed a universal horizon present. In addition, note that the $U(\rho)$ function tends to oscillate, so additional interior universal horizons appear at for example $x \sim -0.7556$ and $x \sim -0.94002$. This is the same type interior oscillation found in \cite{Eling:2006ec,Barausse:2011pu} for asymptotically flat solutions in four-dimensions. To see this behavior more clearly, we plot the function $N_1$ versus $x$ in Figure 3.
\begin{figure}
\begin{center}
 \includegraphics[angle=0,height= 11cm, width=11cm]{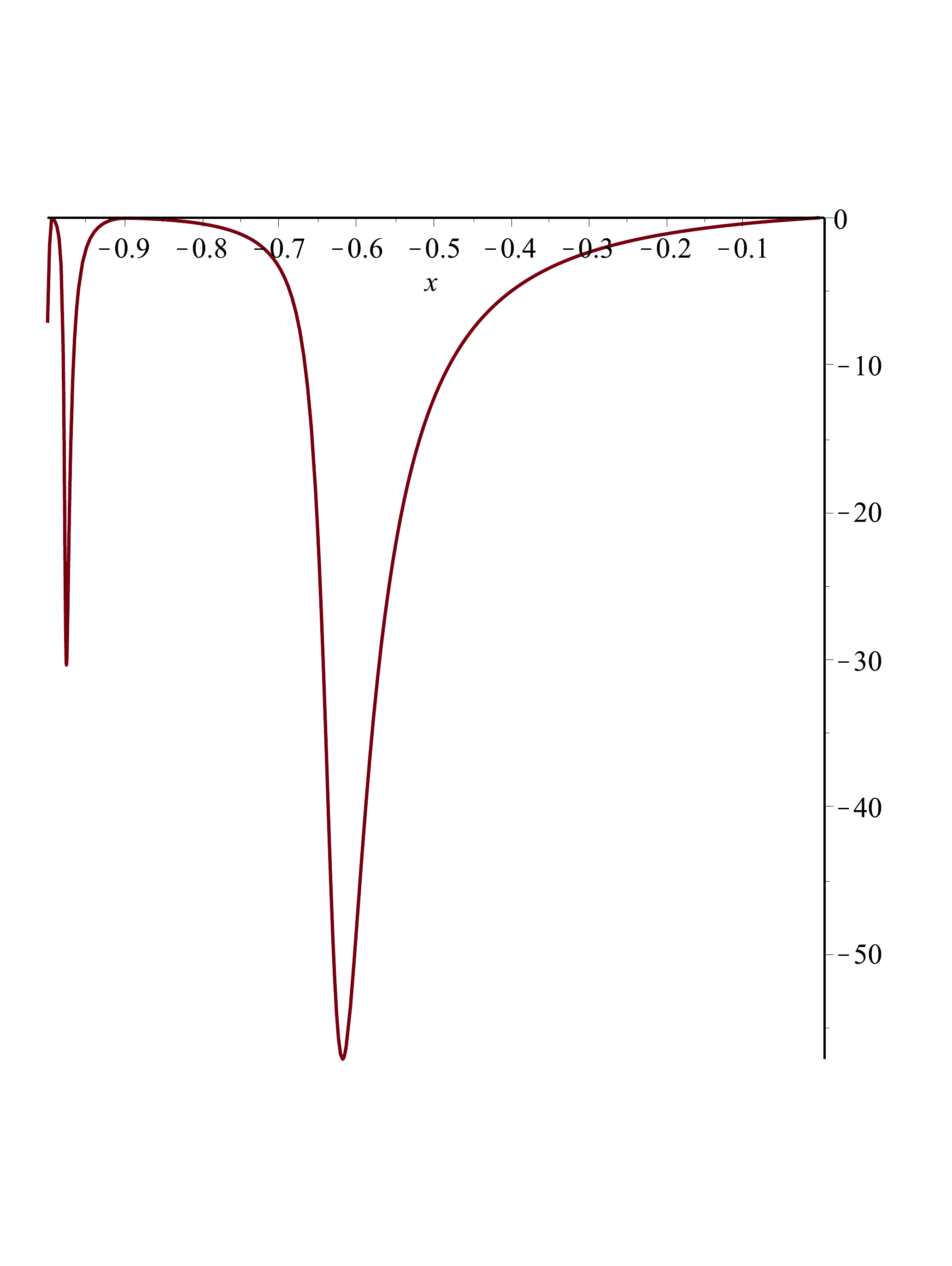}\\
\caption{\label{oscillation}Plot of $N_1$ versus $x$ for the solution in Figure 1. The function begins to oscillate rapidly as the singularity is approached. }
 \end{center}
\end{figure}
Again this is qualitatively the same type of behavior noted previously in the asymptotically flat cases.

The only noticeable change in the functions with $c_3$ is that as its value increases, the size of the interior peak(s) in $K_1$ tends to decrease and the oscillation tends to smooth out. See Figure 4.
\begin{figure}
\begin{center}
 \includegraphics[angle=0,height= 11cm, width=11cm]{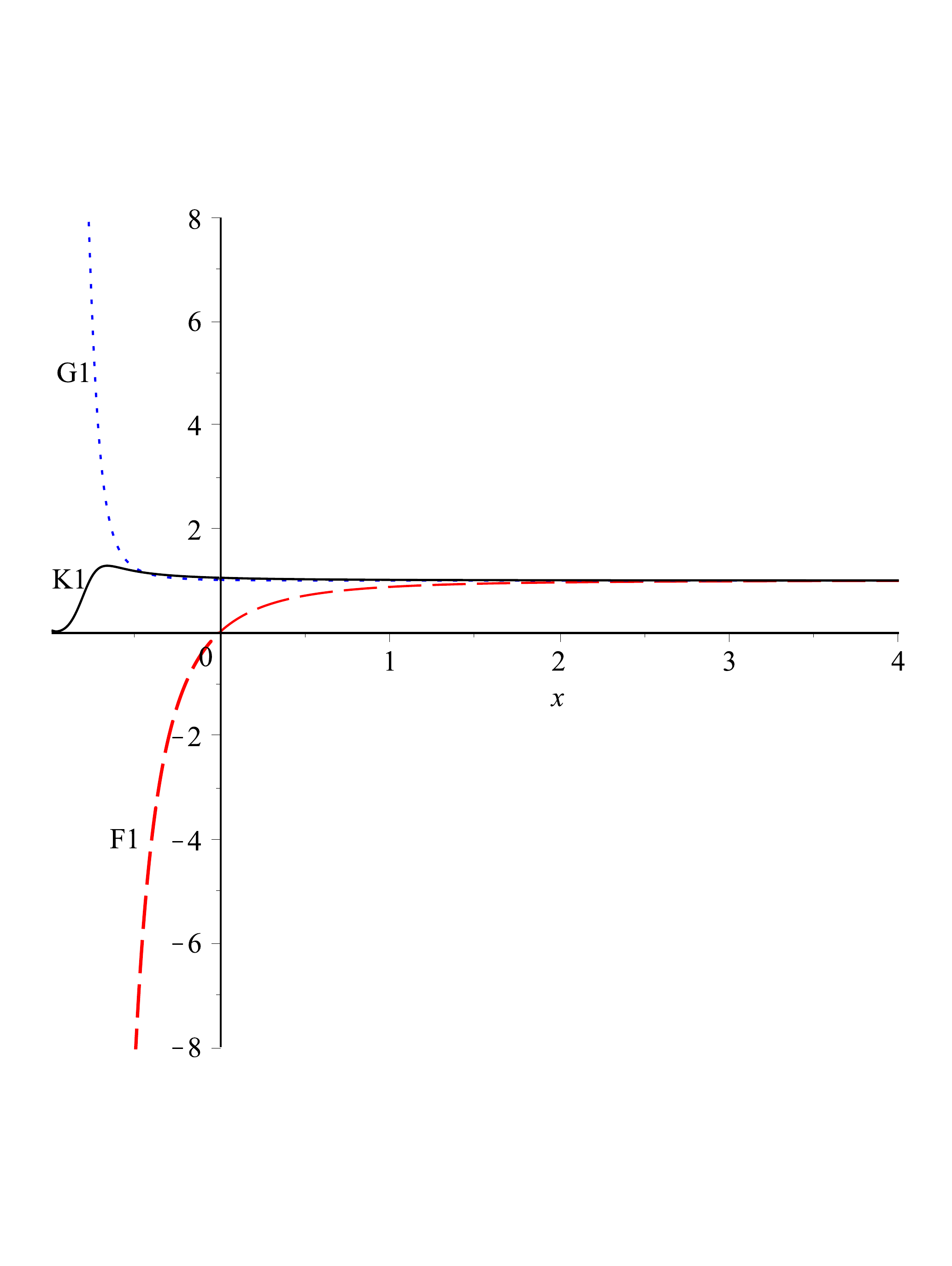}\\
\caption{\label{bhsol1}Plots of $F_1$ (dashed), $G_1$ (dotted) and $K_1$ (solid) vs $x=r-r_0$ for $c_3=0.3$ in $D=2$.}
 \end{center}
\end{figure}
At and beyond $c_3 \sim 0.35$ no solutions appeared to exist generically. However, this particular value range may be an artifact of numerical errors or inaccuracies in our simple first order series solution around the horizon. Solutions may exist for somewhat larger values. Also for larger $c_3$ there may be special solutions present. In \cite{Basu:2016vyz}, a regular black hole solution was found numerically with $c_3 = \frac{9}{10}$, which satisfies (\ref{3dcondition}) for $n_s = 4$. For this choice of $c_3$, we indeed found a nearly regular solution, at a very fined tuned value of the shooting parameter. A singularity appeared at large $x \sim 100$, which seems likely to be due to numerical errors or an inaccuracy.

\subsubsection{$z>2$ Cases- No Generic Solution?}

Now we will consider cases with larger values of the dynamical exponent $z$. First we examined $z=3$. In this case for a range of $c_3$ the behavior of the solutions was different than in $z=2$. As we tuned the value of $K_1(\rho_0)$ the asymptotic values of $N_1$ and $M_1$ tended to remain very close to unity but always below it. For example, for $c_3 =0.1$ for $K_1(\rho_0)$ between 0.94 and 1.75  the value of $N_1$ varied from 0.9843 to 0.9705.  Below $K_1(\rho_0)= 0.94$ the initial conditions become complex. There was never a case where the asymptotic values became greater than unity. We found the same type of behavior for $z=4$ and also $z=5/2$. This indicates there are generically no regular asymptotically Lifshitz black holes for $z>2$, but it would be useful to probe this regime in more detail.

\subsubsection{$z=3/2$ Case}

Perhaps $z=2$ is another special value (like $z=1$) where solutions exist. To check this we considered the case where $z=3/2$. Here we found again regular solutions for a range of $c_i$, similar to the $z=2$ case. In Figure 5, we have plotted the solution $c_3=0.1$, which shows similar behavior to the $z=2$ case (with less oscillation of $K_1$)
\begin{figure}
\begin{center}
 \includegraphics[angle=0,height= 11cm, width=11cm]{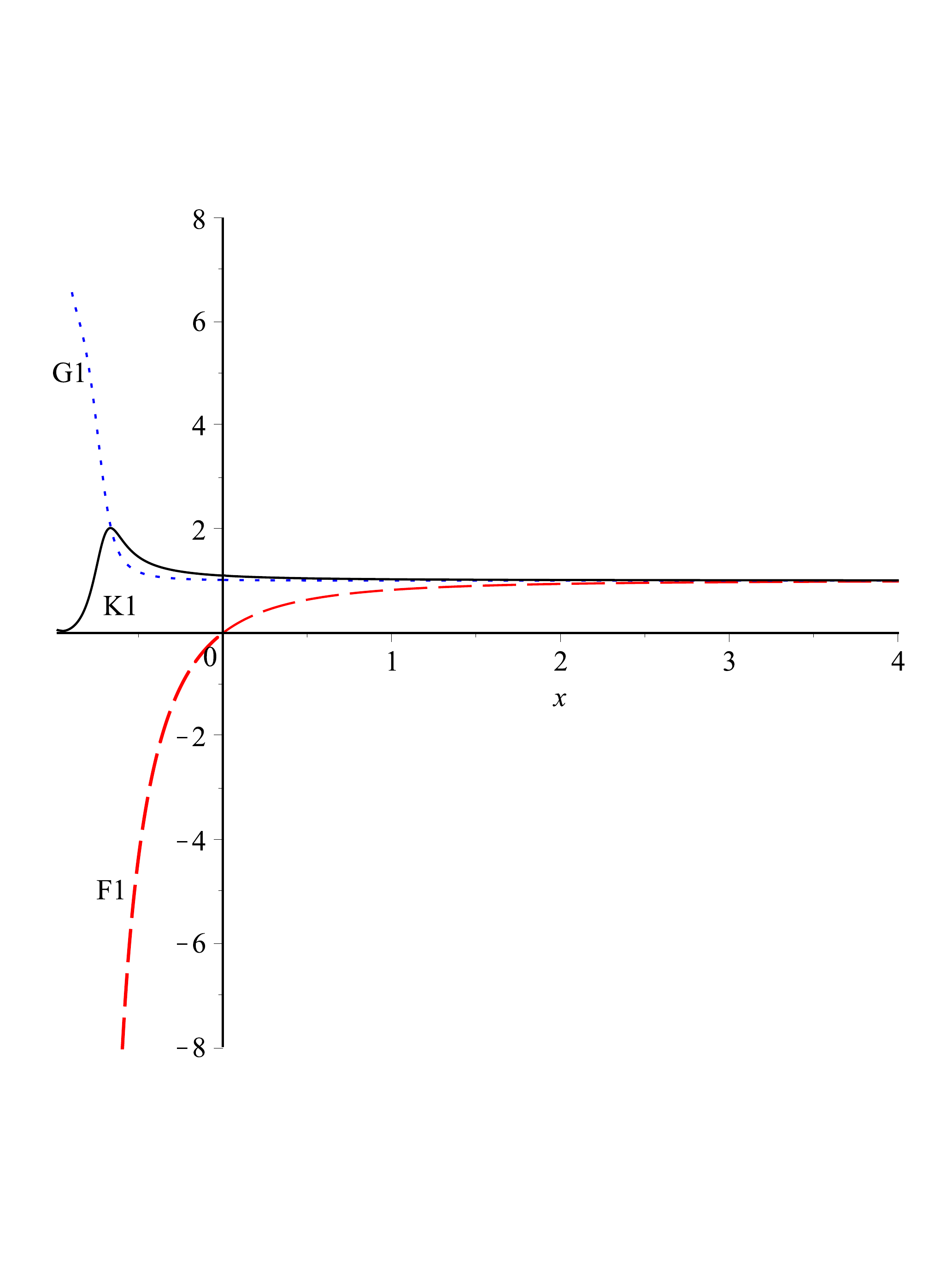}\\
\caption{\label{bhsol1}A $z=3/2$ solution in $D=2$: Plots of $F_1$ (dashed) , $G_1$ (dotted) and $K_1$ (solid) vs $x=r-r_0$ for $c_3=0.1$.  }
 \end{center}
\end{figure}

\subsection{$D=3$ Lifshitz Black Holes}

Now we consider four-dimensional black hole solutions. The behavior in this case throughout is qualitatively similar to the three-dimensional case. First, for a range of $c_3$ (here up to about 0.4) we found regular $z=2$ solutions. An example is shown in Figure 6.

\begin{figure}
\begin{center}
 \includegraphics[angle=0,height= 11cm, width=11cm]{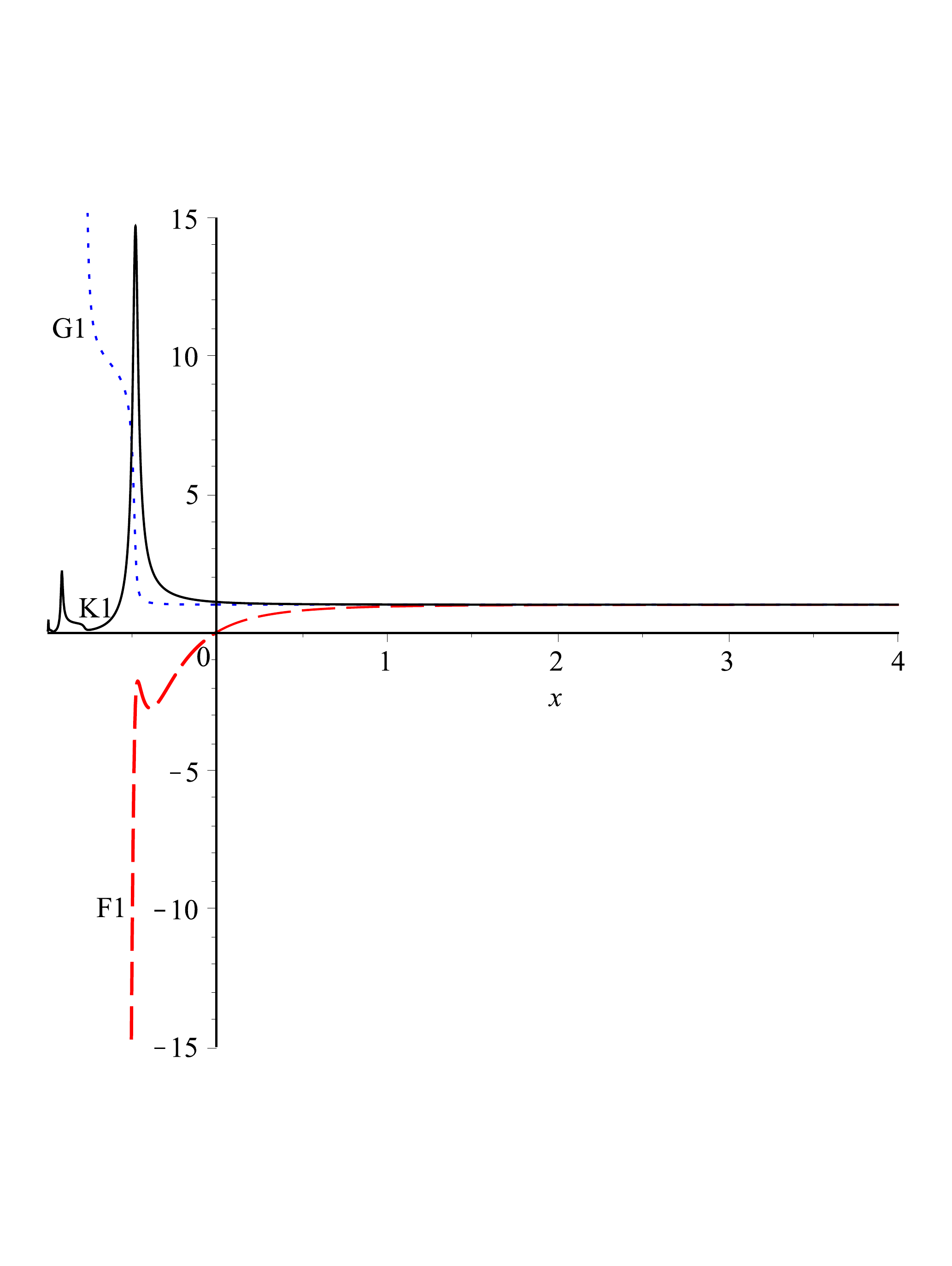}\\
\caption{\label{bhsol1}Plots of $F_1$ (dashed) , $G_1$ (dotted) and $K_1$ (solid) vs $x=r-r_0$ for $D=3$ and $c_3=0.1$. Here $r_0$ is chosen to be 1. }
 \end{center}
\end{figure}
Performing a Non-linear curve fit of our data for $K_1$ to $1-K_{\Delta} \rho^{-\Delta}$ leads to a reasonable agreement between $\Delta$ ($\sim 3.83$) and the value obtained in (\ref{Deltacondition}) ($\sim 3.91$ here) required by the non-analytic expansion around infinity.  In the interior, there are universal horizons associated with the same type of oscillations appearing in $D=2$ oscillations, and as $c_3$ increased the behavior of the functions was the same as in the $D=2$ case.

\subsubsection{$z>2$ Cases- No Solution?}

We again consider $z>2$, following the same methods as in the three-dimensional case.  As we tuned the value of $K1(\rho_0)$ the asymptotic values of $N_1$ and $M_1$ again tended to remain very close to unity but always below it. For example, for $c_3 =0.1$ and for $K1(\rho_0)$ between 1.605  and  0.919 the value of $N_1$ varied from 0.999179 to 0.9995. Below $K_1(\rho_0)= 0.919$ the initial conditions become complex. We again conclude that no asymptotic Lifshitz solutions of this type exist.

\subsubsection{$z=3/2$ Case}

Finally, we again found solutions for $z=3/2$ which are qualitatively similar to the $z=3/2$ solutions in $D=2$. An example is plotted in Figure 7.
\begin{figure}
\begin{center}
 \includegraphics[angle=0,height= 11cm, width=11cm]{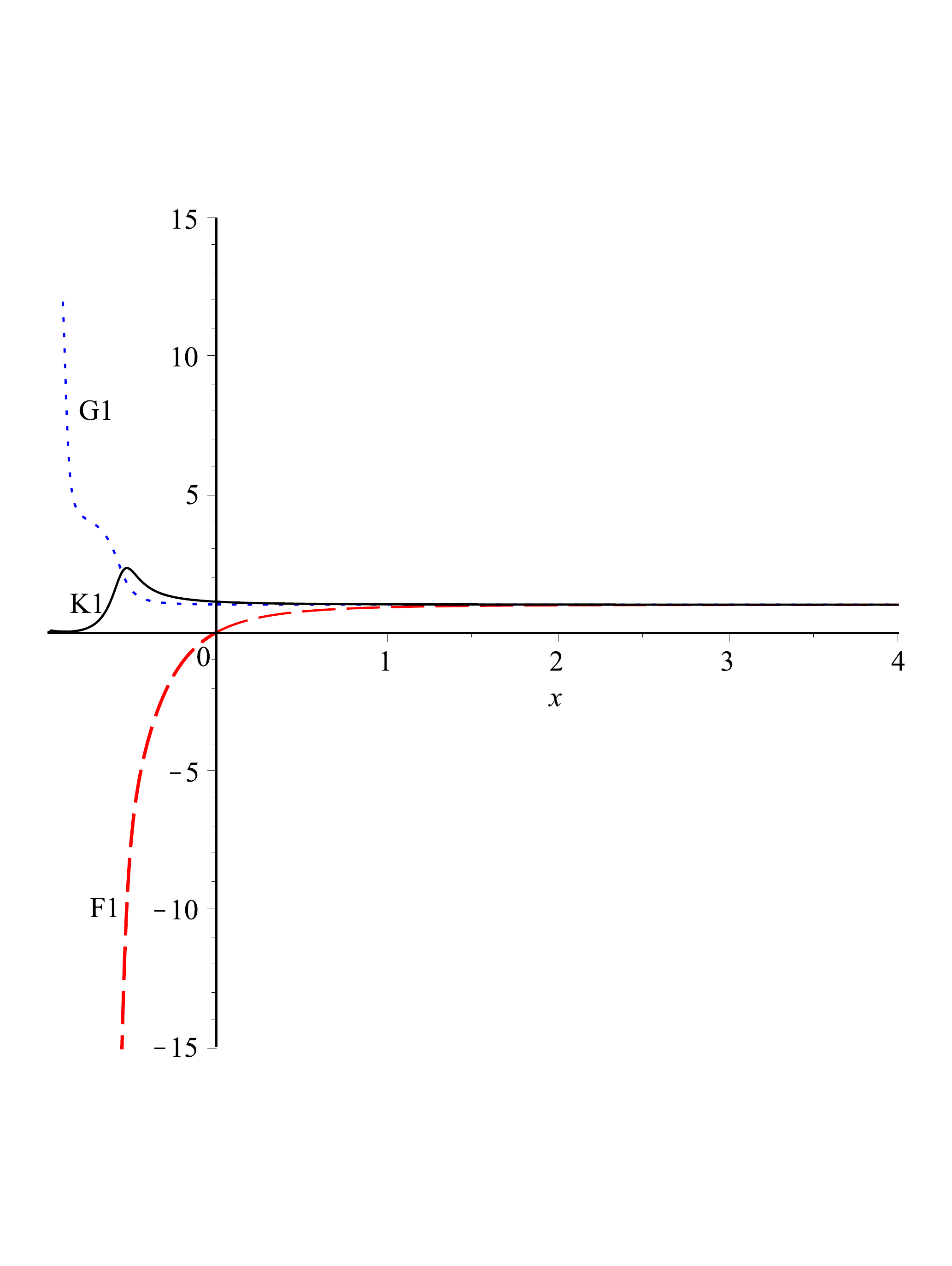}\\
\caption{\label{bhsol1}A $z=3/2$ solution in $D=3$: Plots of $F_1$ (dashed), $G_1$ (dotted) and $K_1$ (solid) vs $x=r-r_0$ for $c_3=0.1$.}
 \end{center}
\end{figure}

\section{Discussion}

We found numerically a new class of $z \neq 1$ asymptotically Lifshitz black hole solutions in three and four dimensional Ho\v{r}ava-Lifshitz gravity. This class of solutions is distinguished by the non-analytic behavior of the solution near infinity. In particular, for $z=2$ and $z= 3/2$, solutions exist for a range of $c_3$. We also studied the properties of the solution inside the horizon and showed there is oscillatory behavior in the aether field, just as in the asymptotically flat case.  Interestingly, for $z>2$, there seem to be generically no solutions of this type, although particular solutions with special values of $c_i$ parameters are likely to exist.

In the future, it would be interesting to understand the nature of the non-analytic asymptotic solution in these cases.  While a regular universal horizon exists, it is not clear whether one can define a corresponding ADM-like mass for these solutions that can used to study black hole thermodynamics, such as the First Law. Are there other distinguishing properties of solutions, such as stability or instability? From the point of view of holography, these black holes describe the thermal state of a dual field theory with $z \neq 1$. What does the nature of the asymptotic solution here tell us about the properties of the dual theory? Finally, for other holographic applications, these solutions (in four-dimensions) may also be useful as background one can perturb in the fluid-gravity setting, following \cite{Eling:2014saa, Davison:2016auk}. For example, it would be interesting use these black holes to study the characteristics of fluid transport coefficients in cases where $z \neq 1$.

\section*{Acknowledgements}
I would like to thank Jishnu Bhattacharyya and David Mattingly for valuable discussions.
This research was supported by the European Research Council
under the European Union's Seventh Framework Programme (ERC Grant
agreement 307955).

\end{document}